\documentclass[12pt]{article}
\usepackage{epsfig}
\usepackage{amssymb}

\newcommand{\lsim}{\buildrel < \over {_\sim}}
\newcommand{\ie}{{\it i.e.}}

\newcommand{\qu}{{\rm q}}

\newcommand{\qbm}{{\rm\bar q}}
\newcommand{\qq}{\qu\qbm}

\newcommand{\pvec}{\vec p}
\newcommand{\kvec}{\vec k}
\newcommand{\rvec}{\vec r}
\newcommand{\Rvec}{\vec R}

\newcommand{\M}{{\cal M}}
\newcommand{\pl}{{||}}
\newcommand{\halft}{\textstyle{\frac{1}{2}}}

\newcommand{\morder}[1]{{\cal O}\left(#1 \right)}
\newcommand{\eq}[1]{(\ref{#1})}

\newcommand{\ave}[1]{\langle{#1}\rangle}
\newcommand{\im}[1]{{\rm Im}{#1}\,}
\newcommand{\re}[1]{{\rm Re}{#1}\,}

\newcommand{\beq}{\begin{equation}}
\newcommand{\eeq}{\end{equation}}
\newcommand{\beqa}{\begin{eqnarray}}
\newcommand{\eeqa}{\end{eqnarray}}
\newcommand{\nn}{\nonumber}

\newcommand{\ov}{\widehat}

\renewcommand{\thefootnote}{\fnsymbol{footnote}}

\begin{document}

\begin{titlepage}
\begin{flushright}
           LAPTH-1098/05\\
           hep-ph/0505066\\
\end{flushright}

\vskip 1.5cm

\centerline{\bf NON-UNIVERSALITY OF TRANSVERSE}
\centerline{\bf COULOMB EXCHANGE AT SMALL {\Large $x$}} 

\vskip 1cm
\centerline{J.~Aichelin$^1$, F.~Arleo$^2$, P.-B.~Gossiaux$^1$,}
\vspace{.2cm}
\centerline{T.~Gousset$^1$, S.~Peign\'e$^{3,1}$, M.~Thomas$^1$}
\vskip .5cm

{\small\sl
\centerline{ $^1$SUBATECH\footnote{UMR 6457, Universit\'e de Nantes,
    Ecole des Mines de Nantes, IN2P3/CNRS.}, 4 rue Alfred Kastler,
  44307 Nantes cedex 3, France}
\vspace{.1cm}
\centerline{ $^2$LPTHE\footnote{Universities Paris VI, VII, and CNRS.},
  4 place Jussieu, 75252 Paris Cedex 05, France}  
\vspace{.1cm}
\centerline{ $^3$LAPTH\footnote{CNRS, UMR 5108, associated to the
      University of Savoie.}, Chemin de Bellevue, 
BP 110, 74941 Annecy-le-Vieux Cedex, France}}

\vskip 1cm
\begin{abstract}  
Within an explicit scalar QED model we compare, at fixed $x \ll 1$,
the leading-twist $K_\perp$-dependent `quark' distribution 
$f_{\qu}(x, K_\perp)$ probed in deep inelastic scattering and
Drell-Yan production, and show that the model is consistent with 
the universality of $f_{\qu}(x, K_\perp)$. The extension of the model 
from the aligned-jet to the 'symmetric' ki\-ne\-ma\-ti\-cal regime 
reveals interesting properties of the physics of Coulomb 
rescatterings when comparing DIS and DY processes. 
At small $x$ the transverse momentum 
$\ave{k_\perp^2}$ induced by multiple scattering on a single centre 
is process dependent, as well as the transverse momentum broadening 
occurring in collisions on a finite size nuclear target. 
\end{abstract}

\end{titlepage}

\newpage
\renewcommand{\thefootnote}{\arabic{footnote}}
\setcounter{footnote}{0}
\setcounter{page}{1}

\section{Introduction}

A significant modification of the quark and gluon distribution
functions in heavy nuclei --~as compared to light targets such as a
proton or deuterium~-- is observed in Deep Inelastic Scattering (DIS)
experiments on nuclei (for a review, see~\cite{arneodo,pillerweise}).
Although such effects manifest themselves on a wide range in the
light-cone momentum fraction $x$ carried by the parton struck in
the target, it is useful to discuss separately two relevant
kinematical regimes at work. In the target rest frame, the typical
lifetime for the hadronic fluctuation of the virtual photon of energy $\nu$ 
is given by the coherence length $l_c = 2\nu/Q^2 \equiv 1/(M\,x)$ 
(where $M$ is the mass of one scattering centre in the target, here a
nucleon in the nucleus). At large $x$ the coherence length remains small compared to the
typical distance $d$ between two centres, $l_c \lesssim d$. This is
the {\it incoherent} regime for which one expects factorization
between the hard production process on a given centre and the
subsequent final state interaction of the hadronic (or partonic)
fluctuation. Conversely, at small $x \ll 1$, the fluctuation scatters 
coherently on several scattering centres. In this regime,
rescatterings actually affect the hard process. Taking the average
distance $d \simeq 2$~fm between two nucleons in heavy nuclei, one
expects the onset of coherence effects such as shadowing below $x
\simeq 0.1$. This is indeed seen in the small $x$ measurements of 
nuclear structure functions performed at CERN and Fermilab by the NMC
and E665 experiments respectively~\cite{nmce665}. In addition to DIS, 
evidence for sha\-do\-wing corrections also comes
from Drell-Yan (DY) data at small $x_2 =  x$ measured by the
fixed-target experiments E772 and E866/NuSea~\cite{e772e866}. More
data are expected at smaller $x$ from the Relativistic Heavy
Ion Collider (RHIC) facility and in a few years from the Large Hadron
Collider (LHC).

Although the nuclear parton densities probed in DIS or through the DY
mechanism appear to be similar within experimental errors, other 
(intrinsically non-perturbative) ob\-ser\-va\-bles 
depend significantly on the considered hard process. In
particular, the transverse momentum nuclear broadening of the DY pair 
\cite{e772e866,na10} is found to be a factor of 5 or more\footnote{Given the various
incident reaction energies and depending on the precise definition
adopted for the nuclear broadening, telling precisely how big the
discrepancy is turns out to be somewhat delicate although it is
statistically significant.} smaller than what is measured in dijet
photo- or hadro-production \cite{e683e609}. Note that a similar discrepancy between 
DY pair and heavy quarkonium transverse momentum broadening 
has also been reported~\cite{e772e866,na10,raufeisen}. 

In the incoherent regime, 
one would expect the broadening for large nuclei to be proportional to
the path length covered by the quark produced in the hard process,
$\ave{p_{\perp}^2} = \Lambda^2 A^{1/3}$, coming from the diffusion in
transverse momentum space due to multiple
scattering~\cite{chiappettapirner}. Moreover the factorization between
the hard process and the rescatterings should make the strength of the
nuclear broadening $\Lambda$ a universal quantity (up to trivial 
geometrical and color factors). Yet the smallness of nuclear broadening
in the DY process seems to contradict universality, which remains so
far not fully understood~\cite{raufeisen,lqs,guo,JKT}. Given the
fact that most of the data lie at the borderline between the
incoherent and coherent regimes, we may wonder whether coherent
effects could be at the origin of this observation.

Since rescattering affects the hard process as soon as $l_c \gtrsim
d$, there is a priori no reason to expect transverse momentum
broadening to be universal at small $x$. However, when comparing
two different processes, it is difficult to foresee in which process
the broadening will be the largest. 

In this paper those questions are addressed by studying in parallel DIS 
and DY within an explicit scalar QED (SQED) model 
in the small $x$ limit. In this limit we will consider two kinematical 
regimes, namely the aligned jet region where the largest component $2\nu$ of the 
incoming light-cone momentum is mostly transferred to 
a single final state particle, and the symmetric region where it 
is transferred to two final state particles. 
In the aligned jet region (where {\it soft} rescatterings contribute to
the total cross section to leading-twist), the model is shown to be consistent 
with the universality of the $K_\perp$-dependent distribution
$f_{{\rm q}/T}(x, K_\perp)$ of the target quark partici\-pa\-ting to the
hard subprocess, for an arbitrary number of scattering
centres in the target, \ie, $f^{DIS}_{{\rm q}/T}(x, K_\perp) = f^{DY}_{{\rm q}/T}(x, K_\perp)$.
In the symmetric region (where soft rescatterings contribute to higher-twist to the cross section), 
the distribution in the Coulomb transfer $k_\perp$ to the outgoing two-particle 
system is different in DIS and DY (for $x \ll 1$).
This non-universality appears both for a pointlike or finite size target. 
In particular, in the symmetric kinematics the nuclear $k_\perp$-broadening is
driven by monopole rescattering in DY production and by dipole rescattering in DIS. 

The paper is organized as follows. In Section~\ref{se:model_lt} we
recall the models of Refs.~\cite{bhmps,peigne} for the leading-twist
quark distributions probed in DIS and DY production on a pointlike
heavy target, which we extend to the case of a finite size target in
Section \ref{sec:nucleartarget}. Section~\ref{sec3} is devoted to the extension
of the model to the symmetric kinematical regime. 
A summary of our results is given in Section~\ref{se:summary}.

\section{Model for leading-twist DIS and DY quark
  distributions: aligned-jet kinematics}\label{se:model_lt} 
\subsection{Single scattering centre}

\vskip 3mm
{\bf Perturbative model for DIS}
\vskip 3mm

The leading-twist $K_\perp$-dependent quark distributions 
$f_{{\rm q}/T}(x,K_\perp)$ in DIS and DY are modelled within 
a scalar QED (SQED) model.
Let us start by briefly recalling the features of the model for
DIS~\cite{bhmps}.  The contribution to the DIS cross section
$\sigma_{DIS}$ (or to the {\it forward} DIS amplitude) studied
in~\cite{bhmps} is obtained by squaring the DIS {\it production}
amplitude shown in Fig.~1a. The target $T$ is chosen to be a scalar
`heavy quark' of momentum $p$ and mass $M$. The incoming virtual
photon of momentum $q$ couples to scalar `light quarks' of mass $m$,
which appear with on-shell momenta $p_1$ and $p_2$ in the final
state. The `electromagnetic' charge of the light quarks is denoted by
$e$ and the light and heavy quarks interact with `strong' coupling
$g$.  We work in a target rest frame where\footnote{We use the
light-cone variables $k^{\pm}=k^0\pm k^z$ to define a momentum
$k=(k^+, k^-, \vec{k}_{\perp})$.} $q=(-Mx_B, q^-, \vec{0}_{\perp})$,
$x_B$ being fixed and $q^- \equiv 2\nu = Q^2/Mx_B \to \infty$ in the
Bjorken limit. In this limit the contribution to the cross section of transverse
virtual photons is subleading (in SQED), and we take the photon to be 
longitudinally polarized in the following.

Studying the effect of Coulomb {\it soft rescatterings} between the light
and heavy quarks {\it at leading-twist}
\footnote{In section 3 we will extend the model to a
higher-twist kinematical domain, where $p_2^-$ scales with $\nu$, \ie, 
the ratio $y=p_2^-/(2\nu)$ is fixed.} requires
concentrating on the aligned-jet region~\cite{pillerweise}, where most
of the photon energy $\nu$ is transferred to the struck quark, \ie, 
$p_1^- \simeq q^- \gg p_2^-$ (see Fig.~1). Moreover, since the hard
scale $\nu$ does not flow in the internal propagators of the lower
part of Fig.~1a, the hard vertex $\gamma^* {\rm q} \to {\rm q}$ is
taken at zeroth order in $g$.  The square of the DIS production
amplitude describes the soft dynamics which can be directly
interpreted (apart from a trivial factor $\propto e^2Q^2$) as a
contribution to the light quark distribution in the target $f_{{\rm
q}/T}(x, K_\perp)$. The momentum $K$ corresponds to the quark probed
in the target and reads $K=k-p_2$, where $k=\sum k_i$ is the total
Coulomb momentum transfer between the light and heavy quarks. In
Fig.~1a we have $K=p_1-q$, and the quark distribution is thus probed at
$x=K^+/p^+\simeq -q^+/p^+ =x_B$. Since $K^+ >0$, it is easy to
realize, for instance in a light-cone time-ordered formulation, that
the hard subprocess in Fig.~1a is indeed $\gamma^* {\rm q} \to {\rm
q}$, as in the infinite momentum frame. We stress that the diagrams
contributing to the $Q^2$ evolution of $f_{\qu/T}$ being excluded, the
model thus describes $f_{\qu/T}$ at an initial soft scale $Q_0$.
 
\begin{figure}[t]
\begin{center}
\epsfig{figure=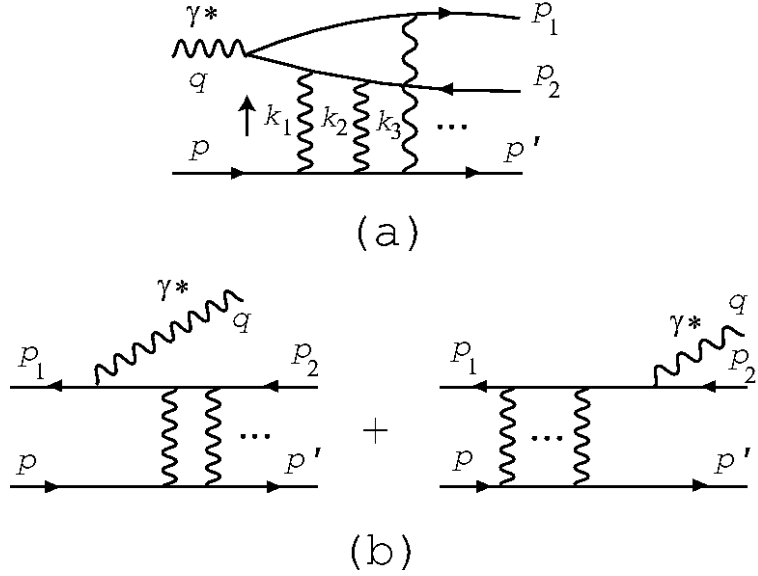,angle=0,width=10cm}
\end{center}
\caption[*]{(a) The DIS production amplitude in the model of 
Ref.~\cite{bhmps}. Coulomb rescatterings are resummed. (b) The DY
production amplitude obtained from (a) by crossing~\cite{peigne}. 
Diagrams where the heavy photon is emitted from an internal quark line 
are suppressed (in covariant gauges) in the limit $x_B \ll 1$.}
\label{fig_feyn_graph}
\end{figure}

The DIS amplitude of Fig.~1a has been calculated 
in~\cite{bhmps} in the limit $x_B \ll 1$. In transverse coordinate
space, the result including any number of Coulomb rescatterings reads 
\beq
\label{MtildeDIS}
\ov{\M}_{DIS}(\rvec_\perp, \Rvec_\perp) 
= \sqrt{4\pi} \, \psi(r_\perp) \, T_{\qq}(\rvec_\perp, \Rvec_\perp) \ \ ,
\eeq
where we define normalized scattering 
amplitudes $\widehat{\M}$ in terms of Feynman amplitudes $\M$ by
\beq
\ov{\M} = \frac{1}{4M\nu} \, \M \ \ .
\label{normamp}
\eeq
In Eq.~\eq{MtildeDIS} the factor $\psi(r_\perp)$ denotes the $\gamma^* \to 
\qu \qbm$ dipole wave function: 
\beq
\psi(r_\perp) = y \sqrt{\alpha} \, Q \, V(m_\pl\, r_\perp) \ \ ,
\eeq
where $\alpha = e^2/(4\pi)$, $y$ is the light-cone
momentum fraction carried away by the `antiquark' $p_2$,  
\beq
y \equiv \frac{p_2^-}{2\nu} \ll 1 \ \ ,
\eeq
and the function $V$ is given by
\beq
V(m_\pl\, r_\perp) \equiv
\int \frac{d^2\pvec_\perp}{(2\pi)^2}
\frac{e^{i\rvec_\perp\cdot\pvec_{\perp}}}{p_\perp^2+m_\pl^2}
= \frac{K_0(m_\pl \,r_\perp)}{2\pi} \, .
\label{Vexpr}
\eeq
Here $r_\perp$ denotes the dipole size and
\beq
m_\pl^2 = p_2^-Mx_B + m^2 = y\,Q^2 + m^2 \ \ .
\label{mpl} 
\eeq

The $\qu \qbm$ dipole scattering amplitude $T_{\qq}$ appearing in 
(\ref{MtildeDIS}) reads
\beq
T_{\qq}(\rvec_\perp, \Rvec_\perp) = -i 
\left(1 - e^{-ig^2 W(\rvec_\perp, \Rvec_\perp)} \right)\ \ , 
\label{dipamp}
\eeq
where $\vec{R}_\perp$ is the impact parameter of the outgoing quark 
and $W$ is the dipole {\it single} scattering amplitude:
\beq
W(\rvec_\perp, \Rvec_\perp) \equiv
\int \frac{d^2\kvec_\perp}{(2\pi)^2}
\frac{1-e^{i\rvec_\perp\cdot\kvec_{\perp}}}{k_\perp^2+\lambda^2}
e^{i\Rvec_\perp\cdot\kvec_{\perp}} = 
\frac{K_0(\lambda R_\perp)-K_0(\lambda |\vec{R}_\perp+\vec{r}_\perp|)}{2\pi} \,.
\label{Wexpr}
\eeq
We have introduced a finite photon mass $\lambda$. Indeed, while the amplitudes $W$ and $T$ are infrared
safe in the $\lambda\to 0$ limit\footnote{We have for instance
$\lim_{\lambda\rightarrow 0} W(\rvec_\perp, \Rvec_\perp)=\frac{1}{2\pi}
\log\left(\frac{|\Rvec_\perp+\rvec_\perp|}{R_\perp} \right)$.}, some other 
quantities are not, such as the dipole scattering cross section $\sigma_{\qq}(r_\perp)$ 
defined below in Eq.~\eq{dipcross}. Since $\sigma_{\qq}(r_\perp)$ is
a basic quantity which will enter our main equations, we choose to 
use the infrared regulator $\lambda$ from now on. This provides 
a mathematically well-defined framework, and will moreover allow to
compare unambiguously DIS with DY production, for which the 
infrared sensitivity appears at the amplitude level (see Eq.~\eq{phase}). 

We stress that
\beq
\label{unitarity}
\left| T_{\qq}(\rvec_\perp, \Rvec_\perp) \right|^2  = -2 
\im{T_{\qq}(\rvec_\perp, \Rvec_\perp)},
\eeq
a unitarity relation which is satisfied by $T_{\qq}$ since the latter
resums Coulomb rescatterings. For later use, we give the
scattering cross section of a $\qu \qbm$ dipole of size $r_\perp$,
\beq
\label{dipcross}
\sigma_{\qq}(r_\perp) = \int d^2 \Rvec_\perp 
\left| T_{\qq}(\rvec_\perp, \Rvec_\perp) \right|^2  =
2i\int d^2 \Rvec_\perp \, T_{\qq}(\rvec_\perp, \Rvec_\perp) \ \  ,
\eeq
where the second equality follows from~\eq{unitarity} and
\beq
\int d^2 \Rvec_\perp  \re{T_{\qq}(\rvec_\perp, \Rvec_\perp)} = 0 \ \ .
\eeq
We give the explicit form of the dipole cross section in our model,
\beq
\label{dipcross2}
\sigma_{\qq}(r_\perp) = 4 \int d^2 \Rvec_\perp \sin^2{\left[\frac{g^2}{4\pi}
\left( K_0(\lambda R_\perp)-K_0(\lambda |\vec{R}_\perp+ \vec{r}_\perp|)
\right) \right]} \, .
\eeq

The DIS production amplitude~\eq{MtildeDIS} incorporates the
leading-twist shadowing effects discussed in~\cite{bhmps}, and
originates from the kinematical domain
\beqa
Q^2, \nu \to \infty \gg p_2^- \gg M \gg k_{i\perp}, p_{i\perp}, k_i^-,
m \gg k_i^+, p_2^+ \sim Mx_B \gg p_1^+ \propto 1/\nu  && \nn \\
p_2^- \ {\rm fixed} \ \Leftrightarrow \ y = \frac{p_2^-}{2\nu}
\to 0 \hskip 4cm &&
\label{scales}
\eeqa
In the first line the first and last inequalities arise from the Bjorken
limit and the aligned-jet kinematics (the latter being stressed in the
second line). The other inequalities arise from the limit
$x_B \ll 1$. All scales other than $\nu$ are soft, \ie, intrinsic to
the target system. Note that taking a relatively large target mass $M$ is not
essential to our analysis but will simplify the expressions of cross
sections. 

Using $d p_2^z/p_2^0 = d p_2^-/p_2^-$, cross sections will be 
given by (we will always assume $y \ll 1$):
\beq
\label{cross}
\frac{d\sigma}{d \log y} = \frac{1}{4\pi} \int
\frac{d^2\vec{p}_{2\perp}}{(2\pi)^2}\,
\frac{d^2\vec{k}_{\perp}}{(2\pi)^2}\,
|\ov{\M}(\pvec_{2\perp},\kvec_\perp)|^2 \ \ ,
\eeq
where the Fourier transform is defined as
\beq
\ov{\M}(\pvec_{2\perp},\kvec_\perp) = \int
d^2\vec{r}_\perp d^2\vec{R}_\perp \,
\ov{\M}(\rvec_\perp, \Rvec_\perp) \, e^{- i\rvec_\perp\cdot\pvec_{2\perp}
- i\Rvec_\perp\cdot\kvec_{\perp}}\ \ .
\label{Fourier}
\eeq

\vskip 3mm
{\bf Model for DY production}
\vskip 3mm

In order to obtain a model for the quark distribution probed in DY
production, the first step is simply to exchange the virtual photon
and the struck quark lines in Fig.~1a, and to replace $q^2=-Q^2 <0$ by
$q^2=Q^2 >0$. The momentum of the DY pair has now $q^+ >0$ and reads
$q=(Mx_B, q^-, \vec{q}_{\perp})$, and the incoming `antiquark' is
chosen with $\vec{p}_{1\perp}=\vec{0}_{\perp}$. We keep the same notation 
for $k$, $p_2$ (see Fig.~\ref{fig_feyn_graph}a), and $K=k-p_2$, implying that 
$\vec K_\perp=\vec q_\perp$ in DY instead of $\vec K_\perp=\vec
p_{1\perp}$ in DIS. In coordinate space the DY production
amplitude pictured in Fig.~1b is simply related to the DIS amplitude
by a phase factor~\cite{peigne},
\beqa
\ov{\M}_{DY}(\vec{r}_{\perp},\vec{R}_{\perp}) &=& 
- \,\,  e^{ig^2 G(R_\perp)} \,\,
\ov{\M}_{DIS}(\vec{r}_{\perp},\vec{R}_{\perp}) 
\label{phase} \\
G(R_\perp) = \int \frac{d^2\vec{k}_\perp}{(2\pi)^2}
\frac{e^{i\vec{R}_\perp\cdot\vec{k}_{\perp}}}{k_\perp^2 + \lambda^2 } 
&=& \frac{K_0(\lambda R_\perp)}{2\pi}\ \ \mathop{\simeq}_{\lambda
  \to 0} \ \  
\frac{1}{2\pi} \log\left(\frac{1}{\lambda R_\perp}\right).
\label{G} 
\eeqa
As already mentioned, in the case of DY production the infrared
sensitivity shows up at the amplitude level 
through the phase shift $g^2G(R_\perp)$. This is a direct consequence of the
fact that the DY production amplitude of  
Fig.~1b involves the scattering of a charge instead of a dipole in
DIS.  

Since the phase factor in~\eq{phase} has no effect on the total DY cross
section, the present model for DIS and DY is consistent with the
universality of the $K_\perp$-integrated quark distribution. In
Ref.~\cite{peigne} it was shown that the non-trivial crossing
\eq{phase} between DIS and DY has nevertheless interesting
consequences. In particular the distribution $d\sigma/d^2\vec{k}_{\perp}$ in the 
Coulomb transfer $\vec{k}_{\perp}$ 
($\vec{k}_{\perp} \neq \vec{K}_{\perp}$) is different in DIS and DY,
\ie, non-universal, as we will recall in section 3.2. 

This result is not in contradiction with the universality 
of the quark distribution $f_{\qu/T}(x, K_\perp)$ 
since in the aligned-jet kinematics \eq{scales} $k_\perp$ is a variable internal 
to the target structure, integrated out in 
$f_{\qu/T}(x, K_\perp)$. Indeed, we show now that $f_{\qu/T}(x, K_\perp)$ 
is universal within the model of Fig.~1, in agreement with factorization 
theorems~\cite{fact1,fact2}. The differential DY cross section at fixed $K_\perp$ is obtained from 
\eq{cross}, 
\beq
\label{dsigma0}
(2\pi)^2 \frac{d\sigma_{DY}}{d \log y \,d^2\vec{K}_\perp} = 
\frac{1}{4\pi} \int \frac{d^2\pvec_{2\perp}}{(2\pi)^2}\, 
|\ov{\M}_{DY}(\pvec_{2\perp},
\vec{k}_{\perp}=\pvec_{2\perp}+\vec{K}_{\perp})|^2 \ \ .
\eeq
Going to transverse coordinate space and using~\eq{phase} we obtain
\beqa
(2\pi)^2 \frac{d\sigma_{DY}}{d \log y \,d^2\vec{K}_\perp} = 
\frac{1}{4\pi} \int d^2\vec{r}_{\perp}
d^2\vec{R}_{\perp}d^2\vec{r}^{\ '}_{\perp}d^2\vec{R}^{\ '}_{\perp}
\delta^2(\vec{r}_{\perp}+\vec{R}_{\perp}-\vec{r}^{\
  '}_{\perp}-\vec{R}^{\ '}_{\perp}) && \nonumber \\
\times \  e^{-i(\vec{R}_{\perp}-\vec{R}^{\ '}_{\perp})\cdot
  \vec{K}_{\perp}} \, e^{ig^2(G(R_{\perp})-G(R^{\ '}_{\perp}))} 
\ov{\M}_{DIS}(\vec{r}_{\perp}, \vec{R}_{\perp})
\ov{\M}_{DIS}^*(\vec{r}^{\ '}_{\perp}, \vec{R}^{\ '}_{\perp})\ \ . && 
\label{dsigma1}
\eeqa
Using the constraint from the delta function, we can rewrite the phase 
difference as
\beqa
G(R_{\perp})-G(R^{\ '}_{\perp}) &=& 
G(R_{\perp})- G(|\vec{R}_{\perp}+\vec{r}_{\perp}|) - 
(G(R^{\ '}_{\perp})- G(|\vec{R}^{\ '}_{\perp}+\vec{r}^{\ '}_{\perp}|))
\nonumber \\ 
&=& W(\vec{r}_{\perp}, \vec{R}_{\perp}) - W(\vec{r}^{\ '}_{\perp}, 
\vec{R}^{\ '}_{\perp})\ \ ,
\label{phasediff1}
\eeqa
since $W(\vec{r}_\perp,\vec{R}_\perp)=G(R_\perp)-
G(|\vec{R}_\perp+\vec{r}_\perp|)$. It is easy to see that
in~\eq{dsigma1}, the phase difference \eq{phasediff1} can  
be absorbed in the expression of the DIS amplitude given by \eq{MtildeDIS} and \eq{dipamp}. 
After the change of variables $r \leftrightarrow r'$, 
$R \leftrightarrow R'$ we obtain
\beqa
(2\pi)^2 \frac{d\sigma_{DY}}{d \log y \,d^2\vec{K}_\perp} &=& 
\frac{1}{4\pi} \int d^2\vec{r}_{\perp}
d^2\vec{R}_{\perp}d^2\vec{r}^{\ '}_{\perp}d^2\vec{R}^{\ '}_{\perp}
\delta^2(\vec{r}_{\perp}+\vec{R}_{\perp}-\vec{r}^{\
  '}_{\perp}-\vec{R}^{\ '}_{\perp}) \nonumber \\
&\times&  e^{-i(\vec{R}^{\ '}_{\perp}-\vec{R}_{\perp})\cdot\vec{K}_{\perp}}
\, \ov{\M}_{DIS}(\vec{r}_{\perp}, \vec{R}_{\perp})
\ov{\M}_{DIS}^*(\vec{r}^{\ '}_{\perp}, \vec{R}^{\ '}_{\perp}) \, .
\label{dsigma2}
\eeqa
Interpreting the differential cross sections as
$f_{\qu/T}(x, \vec{K}_\perp)$ we thus write
\beq
f^{DY}_{\qu/T}(x, \vec{K}_\perp) = f^{DIS}_{\qu/T}(x, -\vec{K}_\perp) 
= f^{DIS}_{\qu/T}(x, \vec{K}_\perp) \ \ ,
\label{univ1}
\eeq
where we used the fact that $f_{\qu/T}(x, \vec{K}_\perp)$ is a
function of $K_\perp \equiv |\vec{K}_\perp|$ only.  We stress that the
universality found in~\eq{univ1} directly translates into observable
quantities. Indeed, in DIS $\vec{K}_\perp = \vec{p}_{1\perp}$
(Fig.~1a) and in DY $\vec{K}_\perp = \vec{q}_{\perp}$ (Fig.~1b). Thus
in the present model the leading-twist DIS struck quark $p_{1\perp}$
distribution and the DY $q_{\perp}$ distribution are
identical. We stress that in DY the target
quark distribution is probed at $x_2= K^+/p^+ \simeq q^+/p^+ = x_B$,
\ie, $x_2 \ll 1$ in our model. As already mentioned in the
Introduction, we expect the {\it coherent} rescattering physics 
to come into play at $x_2 \lsim 0.1$, where nuclear shadowing becomes quantitatively important.

One might ask whether the universality \eq{univ1} would be preserved in a more
realistic model for DY production with a {\it composite} projectile.
The role of spectators in our model for DY production is studied in
Appendix A.  We find that spectator rescatterings do not affect the DY
distribution $d\sigma_{DY}/d^2\vec{q}_\perp$ (see the right hand side of
Eq.~\eq{dsigmadq2} which only depends on the DIS amplitude).  The main
consequence of using a composite projectile is to replace the infrared
cut-off $\lambda$ (the finite photon mass) by the inverse size
$\delta$ of the projectile. We also show that the model with a
composite projectile is consistent with factorization 
(see \eq{convolution} and Ref.~\cite{fact2}).  
Since spectator rescatterings do not lead to any breaking of universality between the
DIS and DY $K_\perp$-distributions (at least in our SQED model), we
will neglect (projectile) spectators in the following, and use for DY
production the model with a pointlike scalar projectile presented
above.

It is worth recalling that the $K_\perp$ distribution\footnote{From
now on we suppress  the subscript `DIS' or `DY' for this universal
distribution.} can be expressed in terms of the $\qu \qbm$ dipole
cross section (see for instance~\cite{KST}). From~\eq{dsigma2} and
\eq{MtildeDIS} we obtain
\beqa
(2\pi)^2 \frac{d\sigma}{d \log y \,d^2\vec{K}_\perp} &=& 
\int d^2\vec{r}_{\perp} d^2\vec{r}^{\ '}_{\perp}
e^{-i(\vec{r}_{\perp}-\vec{r}^{\ '}_{\perp})\cdot \vec{K}_{\perp}}
\,\psi(r_\perp)\, \psi(r^{\ '}_\perp) \nonumber \\
&\times& \int d^2\vec{R}_{\perp} \, T_{\qq}(\vec{r}_{\perp},\vec{R}_{\perp}) 
T_{\qq}^*(\vec{r}^{\ '}_{\perp},\vec{R}_{\perp}+\vec{r}_{\perp}
-\vec{r}^{\ '}_{\perp})  \,.
\label{dsigma3}
\eeqa
With the identity
\beqa
&& T_{\qq}(\vec{r}_{\perp},\vec{R}_{\perp}) 
T_{\qq}^*(\vec{r}^{\ '}_{\perp},\vec{R}_{\perp}+\vec{r}_{\perp}
-\vec{r}^{\ '}_{\perp}) = \nn \\
&& i T_{\qq}(\vec{r}_{\perp},\vec{R}_{\perp}) 
- i T_{\qq}^*(\vec{r}^{\ '}_{\perp},\vec{R}_{\perp}+\vec{r}_{\perp}
-\vec{r}^{\ '}_{\perp})
-i T_{\qq}(\vec{r}_{\perp}-\vec{r}^{\ '}_{\perp}, \vec{R}_{\perp})  
\eeqa
the expression~\eq{dsigma3} becomes (using~\eq{dipcross})
\beqa
(2\pi)^2 \frac{d\sigma}{d \log y \,d^2\vec{K}_\perp} &=& 
\int d^2\vec{r}_{\perp} d^2\vec{r}^{\ '}_{\perp}
e^{-i(\vec{r}_{\perp}-\vec{r}^{\ '}_{\perp})\cdot \vec{K}_{\perp}}
\, \psi(r_\perp) \, \psi(r^{\ '}_\perp) \nn \\
&\ \ \ \times& \left[ \halft \sigma_{\qq}(r_\perp)
+\halft \sigma_{\qq}(r^{\ '}_\perp) - 
\halft  \sigma_{\qq}(|\vec{r}_{\perp}-\vec{r}^{\ '}_{\perp}|) \right]
\, .
\label{dsigmadKsingle}
\eeqa
The latter expression can be found in \cite{KST} (in 
the more general case of finite $y$). 

\subsection{Nuclear target}
\label{sec:nucleartarget}

The model for DIS and DY production on a pointlike target described in
the previous section can be directly generalized to the case of
several scattering centres. Since the DIS and DY amplitudes off a
single centre have been derived in the limit $x_B \ll 1$, they should
also describe the production off a `nuclear target' in the limit where
the nuclear radius $R_A$ is kept smaller than the coherence length
$l_c$,
\beq
\label{fullcoherence}
R_A \ll l_c = \frac{1}{M x_B} \ \ .
\eeq
In this limit the production amplitudes are simply obtained from 
\eq{MtildeDIS} and~\eq{phase} by replacing the scattering potential on
a single centre $G(R_\perp)$ given in~\eq{G} by the scattering
potential on $A$ centres located at transverse\footnote{In the total
coherence limit~\eq{fullcoherence} the longitudinal positions of the
centres are irrelevant and only the thickness function $T$ of the target
enters (see below).} positions $\vec{x}_{i \perp}$,
\beq
G(R_\perp) = \int \frac{d^2\vec{k}_\perp}{(2\pi)^2}
\frac{e^{i\vec{R}_\perp\cdot\vec{k}_{\perp}}}{k_\perp^2 + \lambda^2 } 
\longrightarrow G_A(\vec{R}_\perp) \equiv \sum_{i=1}^A
G(|\vec{R}_\perp - \vec{x}_{i \perp}|) \ \ .
\label{gn}
\eeq
Denoting
\beqa
W_A(\vec{r}_{\perp}, \vec{R}_{\perp}) &=& G_A(\vec{R}_{\perp})-
G_A(\vec{R}_{\perp}+\vec{r}_{\perp}) \label{wngn} \\
T_{\qq}^A(\rvec_\perp, \Rvec_\perp) &=&
-i \left(1 - e^{-ig^2 W_A(\rvec_\perp, \Rvec_\perp)} \right) \, ,
\eeqa
we obtain for the DIS and DY amplitudes on the nuclear target $A$
\beqa
\ov{\M}_{DIS}^A (\rvec_\perp, \Rvec_\perp) 
&=& \sqrt{4\pi} \, \psi(r_\perp) \, 
T_{\qq}^A(\rvec_\perp, \Rvec_\perp) \, ,
\label{MtildeADIS} \\
\ov{\M}_{DY}^A(\vec{r}_{\perp},\vec{R}_{\perp}) &=&
- \,\,  e^{ig^2 G_A(\vec{R}_\perp)} \,\,
\ov{\M}_{DIS}^A(\vec{r}_{\perp},\vec{R}_{\perp}) \,.
\label{MtildeADY}
\eeqa

\begin{figure}[t]
\begin{center}
\epsfig{figure=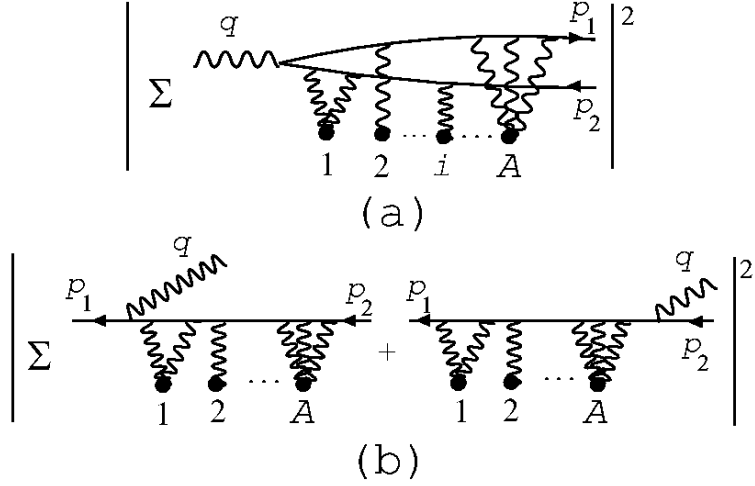,angle=0,width=10cm}
\end{center}
\caption[*]{Model for the DIS (a) and DY (b) nuclear quark distribution 
functions $f^{DIS}_{\qu/A}(x, \vec{K}_\perp)$ and 
$f^{DY}_{\qu/A}(x, \vec{K}_\perp)$. In DIS we have 
$\vec{K}_\perp = \vec{k}_\perp - \vec{p}_{2\perp} = \vec{p}_{1\perp}$
and in the DY case 
$\vec{K}_\perp = \vec{k}_\perp - \vec{p}_{2\perp} = \vec{q}_{\perp}$.
The model takes into account any number of Coulomb rescatterings on
every centre. In the total coherence limit~\eq{fullcoherence} the 
$A$ centres only differ by their relative {\it transverse} positions.}
\label{nuclearfig}
\end{figure}

We can repeat the steps leading from \eq{dsigma0} 
to \eq{univ1} to realize that $\ov{\M}_{DIS}^A$ and $\ov{\M}_{DY}^A$
provide a model for the `nuclear' quark distribution 
function which is consistent with the universality of 
the $K_\perp$-dependent distribution
\beq
f^{DY}_{\qu/A}(x, \vec{K}_\perp) = f^{DIS}_{\qu/A}(x, -\vec{K}_\perp) \
\ .
\label{univA}
\eeq
The physical content of $f^{DIS}_{\qu/A}= f^{DY}_{\qu/A}$ in terms of
Coulomb rescatterings on the $A$ static centres is depicted in 
Fig.~\ref{nuclearfig}.

We now evaluate the universal distribution
$d\sigma^A/d^2\vec{K}_{\perp}$ by averaging over the positions of the
scattering centres. For our purpose it is sufficient to average with a
uniform distribution,
\beq
\label{naverage}
\left<  \ \ \  \right>_A \equiv \int \prod_{i=1}^A
\left( \frac{d^2 \vec{x}_{i\perp}}{S}\right)  \ \ ,
\eeq
where $S = \int d^2 \vec{x}$ is the transverse area of the
target. Inserting~\eq{MtildeADIS} in~\eq{dsigma2} we find
\beqa
(2\pi)^2 \frac{d\sigma^A}{d \log y \,d^2\vec{K}_\perp} = \int
d^2\vec{r}_{\perp} 
d^2\vec{r}^{\ '}_{\perp} 
e^{-i(\vec{r}_{\perp}-\vec{r}^{\ '}_{\perp})\cdot \vec{K}_{\perp}}
\,\psi(r_\perp)\, \psi(r^{\ '}_\perp) \int d^2\vec{R}_{\perp} && \nn \\
\times \ \left< iT_{\qq}^A(\rvec_\perp, \Rvec_\perp) - 
iT_{\qq}^A(\rvec^{\ '}_\perp, \Rvec^{\ '}_\perp)^* - 
iT_{\qq}^A(\rvec_\perp - \rvec^{\ '}_\perp, \Rvec_\perp)
\right>_A \, , &&   
\label{dsigmadKave1}
\eeqa
where $\vec{R}^{\ '}_{\perp} = \vec{R}_{\perp} +\vec{r}_{\perp} -\vec{r}^{\ '}_{\perp}$. 
Using $W_A(\rvec_\perp, \Rvec_\perp) =
\sum_{i=1}^A W(\rvec_\perp, \Rvec_\perp - \vec{x}_{i\perp})$ and the approximation\footnote{With this
approximation and the uniform distribution \eq{naverage}, we can derive Glauber-like expressions 
without resorting to the full formalism of \cite{Glauber}.} 
$\sigma_{\qq}(r_\perp) \ll S$ (valid when the nuclear radius is much 
larger than the interaction range $\lambda^{-1}$), we show using \eq{dipcross} that 
\beq
\left< 1-iT_{\qq}^A(\rvec_\perp, \Rvec_\perp)\right>_A = 0
\eeq
for $\vec{R}_\perp$ outside the nucleus, and 
\beqa
\left< 1-iT_{\qq}^A(\rvec_\perp, \Rvec_\perp)\right>_A = \left<
  e^{-ig^2 W_A(\rvec_\perp, \Rvec_\perp)} \right>_A  
&=& e^{-T \int d^2\vec{x} \, i T_{\qq}(\rvec_\perp, \Rvec_\perp -\vec{x})}
\nn \\ 
&=&  e^{-T \sigma_{\qq}(r_\perp)/2} \, , 
\label{aveexpW}
\eeqa
for $\vec{R}_\perp$ inside the nucleus, where $T=A/S$ is the target `thickness'. 
The average (\ref{naverage}) of the $\qq$ dipole matrix
element $S_{\qq}=1-iT_{\qq}$ resulting in the form (\ref{aveexpW}) allows to interpret the 
rescattering process as a random walk in the
transverse plane (see Ref.~\cite{JKT}). 

The equation~\eq{dsigmadKave1} becomes 
\beqa
(2\pi)^2 \frac{d\sigma^A}{d \log y \,d^2\vec{K}_\perp} = S \int
d^2\vec{r}_{\perp} 
d^2\vec{r}^{\ '}_{\perp}
e^{-i(\vec{r}_{\perp}-\vec{r}^{\ '}_{\perp})\cdot \vec{K}_{\perp}}
\,\psi(r_\perp) \,\psi(r^{\ '}_\perp) \ \ \ \ \ \ &&\nonumber \\
\times \left\{ 1 - e^{-T \sigma_{\qq}(r_\perp)/2} 
- e^{-T \sigma_{\qq}(r^{\ '}_\perp)/2} + 
e^{-T \sigma_{\qq}(|\vec{r}_{\perp}-\vec{r}^{\ '}_{\perp}|)/2}
\right\} \,. && 
\label{dsigmadKave2}
\eeqa 
This corresponds to an `eikonalisation' of the distribution off a
single scattering centre \eq{dsigmadKsingle} (see also~\cite{KST}). In the 
limit of vanishing thickness $T \to 0$, \eq{dsigmadKave2} reproduces 
\eq{dsigmadKsingle} up to the factor
$A$. Since $d\sigma^A/d^2\vec{K}_\perp$ is universal, we see that the
DY $q_\perp$-distribution (and a fortiori the total DY cross section)
can be expressed in terms of the $\qu \qbm$ dipole scattering cross
section, as is the case for DIS~\cite{nikzak}. 

\section{Extension of the model to the symmetric kinematical regime}
\label{sec3}

\subsection{Symmetric kinematics and interpretation}

In the preceding section we have focused on leading-twist
contributions to the DIS and DY cross sections, arising from the
aligned-jet region defined in~\eq{scales}. We would like to stress
however that the validity of our calculations so far is not restricted to 
the aligned-jet domain. Since we only used $p_2^- \ll \nu$, they remain
correct even when $p_2^-$ scales with $\nu$ (\ie, in the `symmetric'
kinematical region), provided the ratio $y = p_2^-/(2\nu)$ is fixed to
a small finite value, \ie, in the region
\beqa
&& \hskip -5mm p_1^- \sim \nu \to \infty \gg p_2^- \gg M \gg k_{i\perp},
p_{i\perp}, k_i^-, m \gg 
k_i^+ \sim Mx_B \gg p_1^+, p_2^+ \propto 1/\nu \nonumber \\
&& \hskip 2cm {\rm fixed}\  y = \frac{p_2^-}{2\nu}  \ll 1 \
\Leftrightarrow \ p_2^- \ {\rm scales \ as} \ \nu \ \ .
\label{symmetric}
\eeqa
In fact our results can easily be extended to the domain where 
$y$ is not small as compared to $1$. 
For simplicity we concentrate on the $y \ll 1$ limit, 
but expect our following considerations to apply also to situations
where $y \sim 1$, such as quarkonium ($y \simeq 1/2$) and dijet
leptoproduction.

With the symmetric kinematics~\eq{symmetric} the  interpretation of
the contributions to the DIS and DY cross sections we evaluated is
modified.
\begin{itemize}
\item[(i)] The antiquark of final momentum $p_2$ is now part of the hard
subprocess, which reads $\gamma^* {\rm g} \rightarrow \qu \qbm$ in DIS
(Fig.~1a) and $\qbm {\rm g} \rightarrow \qbm \gamma^*$ in DY
(Fig.~1b).  The DIS and DY processes are now interpreted respectively
as dijet leptoproduction and the associated production of a DY pair
and a jet. In the symmetric kinematics these processes probe the 
$k_\perp$-dependent `gluon' distribution $f_{{\rm g}/T}(x_B, \vec{k}_{\perp})$. 
\item[(ii)] The transverse momentum transfer to the hard system now
corresponds to the Coulomb $k_\perp$ exchange, \ie, not to
$\vec{K}_\perp=\vec{k}_\perp-\vec{p}_{2\perp}$ any longer. In the
case of DIS, $\vec{k}_\perp = \vec{p}_{1\perp} + \vec{p}_{2\perp}$ is
the momentum imbalance between the jets. In DY, 
$\vec{k}_\perp = \vec{q}_{\perp} + \vec{p}_{2\perp}$ is the imbalance between the DY 
pair and the associated produced jet. 
\end{itemize}

In the following we concentrate, in the region~\eq{symmetric}, on the
DIS and DY $k_\perp$ distributions, and show that those are different.

In section 3.2 we show that the $k_\perp$-distribution on a single scattering
centre becomes non-universal beyond leading order in $g^2$. 
In particular higher order corrections to 
$d\sigma_{DY}/d^2\vec{k}_\perp$ vanish in the $\lambda \to 0$ limit 
(at fixed $k_\perp$), contrary to DIS. In section
3.3 we use the model for a nuclear target presented in section 2.2 
to investigate the target size dependence of $d\sigma^A/d^2
\vec{k}_\perp$, and find 
that the strength of $k_\perp$-broadening depends on the process.

\subsection{Production off a single centre: non-universality of the
  $k_\perp$-dependent distribution}  

In this section we show that in the symmetric kinematics
\eq{symmetric} the DIS (dijet) and DY (+ jet) 
$k_\perp$-distributions are different.
We first consider the DIS model presented in section 2.1. From 
\eq{cross} the differential DIS cross section {\it at fixed} $k_\perp$
reads
\beqa
(2\pi)^2 \frac{d\sigma_{DIS}}{d \log y \,d^2\vec{k}_\perp} &=&
\frac{1}{4\pi}\int 
\frac{d^2\pvec_{2\perp}}{(2\pi)^2}\, 
|\ov{\M}_{DIS}(\pvec_{2\perp}, \vec{k}_{\perp})|^2 \nn \\
&=&  \frac{1}{4\pi} \int
d^2\vec{r}_{\perp} |\ov{\M}_{DIS}(\vec{r}_{\perp}, \vec{k}_{\perp})|^2
\, ,
\label{dsigmaDISdkt}
\eeqa
where 
\beq
\ov{\M}_{DIS}(\vec{r}_{\perp}, \vec{k}_{\perp}) 
= \int d^2\vec{R}_{\perp} \, e^{-i \vec{R}_{\perp} \cdot \vec{k}_{\perp}}
\, \ov{\M}_{DIS}(\vec{r}_{\perp}, \vec{R}_{\perp}) \, .
\label{mixed}
\eeq
From \eq{MtildeDIS} we obtain
\beq
\frac{d\sigma_{DIS}}{d \log y \,d^2\vec{k}_\perp} = 
\int d^2\vec{r}_{\perp}\, |\psi(r_\perp)|^2 \,
\frac{d\sigma_{\qq}(\vec{r}_\perp,\vec{k}_\perp)}{d^2\vec{k}_\perp}
\label{dsigmaDISdksingle} \, , 
\eeq
where we defined
\beq
\frac{d\sigma_{\qq}(\vec{r}_{\perp},\vec{k}_{\perp})}{d^2
  \vec{k}_{\perp}} \equiv  
\frac{|T_{\qq}(\vec{r}_{\perp},\vec{k}_{\perp})|^2}{(2\pi)^2} \,  .
\label{dsigmardk}
\eeq

Let us turn to the case of the DY process, 
\beq
(2\pi)^2 \frac{d\sigma_{DY}}{d \log y \,d^2\vec{k}_\perp} = 
\frac{1}{4\pi} \int
d^2\vec{r}_{\perp} |\ov{\M}_{DY}(\vec{r}_{\perp}, \vec{k}_{\perp})|^2
\label{dsigmaDYdkt0} \, , 
\eeq
where $\ov{\M}_{DY}(\vec{r}_{\perp}, \vec{k}_{\perp})$ can be 
shown using \eq{MtildeDIS} and \eq{phase} to be proportional to its
own Born value, 
\beqa
\ov{\M}_{DY}(\vec{r}_{\perp}, \vec{k}_{\perp}) &=& 
- \int d^2\vec{R}_{\perp}\,e^{-i \vec{R}_{\perp} \cdot
\vec{k}_{\perp}} \,  e^{ig^2 G(R_\perp)} 
\ov{\M}_{DIS}(\vec{r}_{\perp},\vec{R}_{\perp}) \label{MDYmixed} \\
&=& C(k_\perp^2) \, 
\ov{\M}_{DY}^{{\rm Born}}(\vec{r}_{\perp}, \vec{k}_{\perp})  \, ,
\label{Cprop}
\eeqa
where we define
\beqa
C(k_\perp^2) &=& \frac{k_\perp^2+\lambda^2}{ig^2} \int d^2\vec{R}_{\perp}\,
e^{ig^2 G(R_\perp)} \,e^{-i \vec{R}_{\perp}\cdot\vec{k}_{\perp}} 
\label{Cfactor} \\
\ov{\M}_{DY}^{\rm Born}(\vec{r}_{\perp}, \vec{k}_{\perp}) &=& 
- \sqrt{4\pi} \, \psi(r_\perp) \, T_{\qq}^{\rm
  Born}(\vec{r}_\perp,\vec{k}_{\perp}) \label{MDYborn} \\ 
T_{\qq}^{\rm Born}(\rvec_\perp, \vec{k}_{\perp})&=& -2i g^2 
\frac{\sin{(\rvec_\perp \cdot \vec{k}_{\perp}/2 )}}{k_\perp^2+ \lambda^2} \, 
e^{i \rvec_\perp \cdot \vec{k}_{\perp} /2} \,. \label{Tqqborn}
\eeqa
Using \eq{Cprop}, \eq{MDYborn} and \eq{dsigmardk} we find 
\beq
\frac{d\sigma_{DY}}{d \log y \,d^2\vec{k}_\perp} = |C(k_\perp^2)|^2 
\int d^2\vec{r}_{\perp}\, |\psi(r_\perp)|^2 \,
\frac{d\sigma_{\qq}^{\rm
    Born}(\vec{r}_\perp,\vec{k}_\perp)}{d^2\vec{k}_\perp}
\label{dsigmaDYdk1} \, .
\eeq
This can be reexpressed as
\beq
(2\pi)^2 \frac{d\sigma_{DY}}{d \log y \,d^2\vec{k}_\perp} = 
\int d^2\vec{r}_{\perp}\, |\psi(r_\perp)|^2 \, 4
\sin^2(\frac{\rvec_\perp \cdot \vec{k}_{\perp}}{2}) \int
d^2\vec{b}\,   e^{i \vec{b}\cdot 
    \vec{k}_{\perp}}\,  \left[-\halft \sigma_{\qq}(b) \right] 
\label{dsigmaDYdk2}
\eeq
by using \eq{Cfactor} and the identity
\beq
\left| \int d^2\vec{R}_{\perp}\,
e^{ig^2 G(R_\perp)} \,e^{-i
    \vec{R}_{\perp}\cdot\vec{k}_{\perp}} \right|^2 = \int d^2\vec{b}\,
e^{i \vec{b}\cdot \vec{k}_{\perp}} \int d^2\vec{R}_{\perp}\,
(i T_{\qq}^{*}(\vec{b}, \vec{R}_{\perp}) +1) \, .
\eeq
Using \eq{dsigmaDYdk1} and \eq{dsigmaDYdk2} we also obtain
\beq
|C(k_\perp^2)|^2 = \frac{\int d^2\vec{b}\, e^{i \vec{b}\cdot \vec{k}_{\perp}}\,
  \sigma_{\qq}(b)}{\int d^2\vec{b}\, e^{i \vec{b}\cdot \vec{k}_{\perp}}\,
  \sigma_{\qq}^{\rm Born}(b)} 
\label{dsigmaDYdk3} \, .
\eeq
Comparing \eq{dsigmaDISdksingle} and \eq{dsigmaDYdk1} it is clear that 
the $k_\perp$-distribution is process dependent. 
We can stress this point by noting that for $k_\perp \gg \lambda$,
the DY distribution \eq{dsigmaDYdk1} equals the Born 
distribution, contrary to the DIS distribution. 
Let us consider the $k_\perp \to \infty$ limit 
at fixed $\lambda$. In the expression of $|C(k_\perp^2)|^2$ given 
in \eq{dsigmaDYdk3}, the phase factor  
$e^{i \vec{b}\cdot \vec{k}_{\perp}}$ rapidly oscillates, except if 
$b \lsim 1/k_\perp \to 0$. When $b \to 0$ the dipole cross section 
can be approximated by its Born value (see \eq{dipcross2}). Thus when 
$k_\perp \gg \lambda$ we have $|C(k_\perp^2)|^2 \simeq 1$, yielding
\beq
\left. \frac{d\sigma_{DY}}{d \log y \,d^2\vec{k}_\perp}
\right|_{k_\perp \gg \lambda}=  
\int d^2\vec{r}_{\perp}\, |\psi(r_\perp)|^2 \,
\frac{d\sigma_{\qq}^{\rm
    Born}(\vec{r}_\perp,\vec{k}_\perp)}{d^2\vec{k}_\perp}
= \frac{d\sigma_{DY}^{\rm Born}}{d \log y \,d^2\vec{k}_\perp}
\label{dsigmaDYdklimit} \, .
\eeq
We can obtain this latter equation more rigorously by using
\eq{dsigmaDYdk1} and calculating $C(k_\perp^2)$ given in \eq{Cfactor} 
for $\lambda \to 0$, the other scales being fixed.  
Using $G(R_\perp)\simeq -\log(\lambda R_\perp)/(2\pi)$ we get 
\beq
C(k_\perp^2) \mathop{\sim}_{\lambda \to 0} \ \  -
\frac{\Gamma(-\frac{ig^2}{4\pi})}{\Gamma(\frac{ig^2}{4\pi})} \, 
e^{\frac{ig^2}{2\pi} \left(\log\left(\frac{k_\perp}{\lambda}\right)
-\gamma\right)} \, \left[ 1 + \morder{\lambda}  \right]
\Rightarrow |C(k_\perp^2)|=1 \, .
\label{C}
\eeq
Thus $C(k_\perp^2 \gg \lambda^2)$ is a pure phase factor, 
leading to \eq{dsigmaDYdklimit}, which confirms to all orders the result 
shown in~\cite{peigne} at next-to-leading order in $g^2$. 

Since the {\it $k_\perp$-integrated} DIS and DY cross sections are
identical (and different from their Born value), the result
\eq{dsigmaDYdklimit} implies that in the $\lambda \to 0$ limit and in the DY case, 
only vanishing $k_\perp \sim \lambda \to 0$ contributes to $\Delta
\sigma \equiv \sigma^{\rm tot} - \sigma^{\rm Born}$. 
This difference observed between the 
$k_\perp$-dependent DIS and DY cross sections is similar to what Bethe
and Maximon found in the case of high energy pair production and
bremsstrahlung~\cite{bm}. In the present context the effect clearly
arises from the infrared divergent Coulomb phase 
(for $\lambda \to 0$) in the DY production amplitude, and suggests that in 
collisions on a proton, the $k_\perp$ exchange in DY + jet production might be 
smaller than in dijet leptoproduction, and thus not `intrinsic' to the target. 

In a realistic situation, when the projectile and target are composite
and neutral, the DY production amplitude is infrared
finite. Effectively, the role of the infrared cut-off $\lambda$ is
played by the {\it largest} infrared momentum cut-off at disposal,
given by the inverse size $\delta$ of the smallest incoming hadron,
projectile or target.  We expect the typical transverse momentum 
contributing to $\Delta \sigma_{DY}$ to be $k_\perp^2 \sim \delta^{2}$ 
instead of $k_\perp^2 \sim \lambda^2$, as discussed in the end of Appendix A.

In the next section, we investigate the non-universality of the 
$k_\perp$-distribution in the case of a nuclear target.

\subsection{Nuclear target: non-universality of $k_\perp$-broadening}

The `nuclear target' model of section \ref{sec:nucleartarget} is now
used to derive the target size dependence of the DIS and DY
$k_\perp$-distributions, where $k_\perp$ is the total transverse
Coulomb exchange. We have shown that the distribution in 
$\vec{K}_\perp = \vec{k}_\perp - \vec{p}_{2\perp}$ is
universal (see~\eq{univA}), and this holds in the two kinematical domains 
\eq{scales} and \eq{symmetric}.

On the contrary, we now explicitly show that the distribution 
in the transverse Coulomb exchange $d\sigma^A/d^2 \vec{k}_\perp$ 
is process dependent. Similarly to the case of a single 
scattering centre, this can easily be guessed from
Eq.~\eq{MtildeADY}. The DIS and DY amplitudes differ by a pure phase
factor in $R_\perp$ space, but there is no reason to expect this
to hold in the space of the conjugate variable $k_\perp$. 

Inserting~\eq{MtildeADIS} in~\eq{dsigmaDISdkt} we get
\beq
\frac{d\sigma^A_{DIS}}{d \log y \,d^2\vec{k}_\perp} = 
\int d^2\vec{r}_{\perp}\, |\psi(r_\perp)|^2 \,
\frac{d\sigma^A_{\qq}(\vec{r}_\perp,\vec{k}_\perp)}{d^2\vec{k}_\perp}
\label{dsigmaDISdkstruct} \, , 
\eeq
with
\beqa
\frac{d\sigma^A_{\qq}}{d^2\vec{k}_\perp}
&=& \frac{1}{(2\pi)^2} \int d^2\vec{R}_{\perp}
d^2\vec{R}^{\ '}_{\perp}
e^{-i(\vec{R}_{\perp}-\vec{R}^{\ '}_{\perp})
\cdot \vec{k}_{\perp}}\times\\
&&\!\!\! \left< 1- e^{-ig^2 W_A(\rvec_\perp, \Rvec_\perp)} -
e^{ig^2 W_A(\rvec_\perp, \Rvec^{\ '}_\perp)}
+ e^{-ig^2 (W_A(\rvec_\perp, \Rvec_\perp)
- W_A(\rvec_\perp, \Rvec^{\ '}_\perp))} \right>_A 
\nonumber
\eeqa
representing the differential elastic scattering cross section of a $\qq$ dipole of 
size $r_\perp$ on a nuclear target. 
The form \eq{dsigmaDISdkstruct} emphasizes the decoupling 
between production and rescattering of the $\qq$ dipole in
the process. The average over the positions of the scattering centres
leads to
\beqa
\label{dsigmaqqdk}
\frac{d\sigma^A_{\qq}}{d^2\vec{k}_\perp} = \frac{S}{(2\pi)^2}
 \int d^2\vec{b} \, e^{i\vec{b}\cdot \vec{k}_{\perp}}
&& \hskip -5mm \left\{ 1 - 2 e^{-T \sigma_{\qq}(r_\perp)/2\phantom{\int} } \right. \nn \\ && + \left. 
e^{-T\sigma_{\qq}(r_\perp)} e^{T \int d^2\vec{x}\,
T_{\qq}(\vec{r}_{\perp},\vec{x})
T_{\qq}^*(\vec{r}_{\perp},\vec{x}+\vec{b})} \right\}
\label{dsigmaDISdkave1}
\eeqa
Using \eq{dsigmardk} we eventually obtain
\beqa
(2\pi)^2 \frac{d\sigma^A_{DIS}}{d \log y \, d^2\vec{k}_\perp} &=& 
S \int d^2\vec{r}_{\perp} \, |\psi(r_\perp)|^2 
\int d^2\vec{b}\,  e^{i \vec{b}\cdot \vec{k}_{\perp}} \nonumber \\
&& \left\{ 1 - 2 e^{-T\sigma_{\qq}(r_\perp)/2} + 
e^{- T \int d^2 \vec{l}\,(1- e^{-i \vec{b} \cdot \vec{l}}) 
\frac{d\sigma_{\qq}(\vec{r}_{\perp},\vec{l})}{d^2 \vec{l}}} \right\}
\, .
\label{dsigmaDISdkave2}
\eeqa

A similar calculation for DY production is performed
inserting the Fourier transform of~\eq{MtildeADY} into~\eq{dsigmaDYdkt0},
\begin{eqnarray}
&& (2\pi)^2 \frac{d\sigma^A_{DY}}{d \log y\,d^2\vec{k}_\perp} = \nn \\
&=& \int d^2\vec{r}_\perp |\psi(r_\perp)|^2 \left<\left|\int d^2\vec{R}_\perp
\left(e^{i g^2 G_A(R_\perp)}-e^{i g^2 G_A(|\vec{R}_\perp+
\vec{r}_\perp|)}\right)e^{-i \vec{R}_\perp\cdot\vec{k}_\perp}\,
\right|^2\right>_A \nonumber\\
&=&
\int\!\!\!d^2\vec{r}_\perp |\psi(r_\perp)|^2 
\left\langle\left|(1-e^{i\vec{k}_\perp\cdot\vec{r}_\perp})
\int\!\!\!d^2\vec{R}_\perp e^{i g^2 G_A(R_\perp)-i 
\vec{R}_\perp\cdot\vec{k}_\perp}\right|^2\right\rangle_A 
\end{eqnarray}
giving
\begin{equation}
\frac{d\sigma^A_{DY}}{d \log y\,d^2\vec{k}_\perp} = \frac{d\sigma_\qu^A}{d^2\vec{k}_\perp}\, 
\int d^2\vec{r}_\perp |\psi(r_\perp)|^2  4 \sin^2(\frac{\vec{r}_\perp\cdot
\vec{k}_\perp}{2}) \,,
\label{dsigmaDYdkstruct}
\end{equation}
where ${d\sigma_\qu^A}/{d^2\vec{k}_\perp}$ is the differential
elastic scattering cross section 
of a quark on a nuclear target in our model, \ie\footnote{We neglect 
contributions $\sim \delta^{(2)}(\vec{k}_\perp)$ which 
do not contribute to ${d\sigma^A_{DY}}/{d^2\vec{k}_\perp}$
due to the factor $\sin^2(\vec{r}_\perp\cdot \vec{k}_\perp /2)$ in
\eq{dsigmaDYdkstruct}.}:  
\begin{equation}
\left. \frac{d\sigma_\qu^A}{d^2\vec{k}_\perp} \right|_{k_\perp \neq
  0}= \frac{1}{(2\pi)^2} 
\int d^2\vec{R}_{\perp}d^2\vec{R}^{\ '}_{\perp}
e^{-i(\vec{R}_{\perp}-\vec{R}^{\ '}_{\perp})\cdot \vec{k}_{\perp}}
\left< e^{-ig^2 \left(G_A(R'_\perp)-G_A(R_\perp)\right)}\right>_A\,.
\end{equation}
We perform the average using $G_A(R'_\perp)-G_A(R_\perp)=
W_A(\vec{R}_\perp-\vec{R}_\perp',\vec{R}_\perp')$ and (\ref{aveexpW}):
\beq
\left. \frac{d\sigma^A_{\qu}}{d^2\vec{k}_\perp}\right|_{k_\perp \neq
  0} = \frac{S}{(2\pi)^2} 
\int d^2\vec{b}\, e^{i \vec{b}\cdot \vec{k}_{\perp}}\, 
e^{-T\sigma_{\qq}(b)/2}\,,
\label{dsigmamon}
\eeq
finally leading to
\begin{equation}
(2\pi)^2 \frac{d\sigma^A_{DY}}{d \log y\,d^2\vec{k}_\perp} = 
S \int d^2\vec{r}_\perp |\psi(r_\perp)|^2 4
\sin^2{(\frac{\vec{r}_\perp\cdot \vec{k}_\perp}{2})}
\int d^2\vec{b}\, e^{i \vec{b}\cdot \vec{k}_{\perp}}\, 
e^{-T\sigma_{\qq}(b)/2}\,.
\label{dsigmaDYdkave2}
\end{equation} 

In line with the above discussion for DIS, (\ref{dsigmaDYdkave2})
exhibits some decoupling between production and rescattering, up to the factor 
$4 \sin^2{(\vec{r}_\perp\cdot \vec{k}_\perp/2)}$ specific to 
DY production. Notice that despite the formal appearance of $\sigma_{\qq}(b)$ in its expression \eq{dsigmamon}, 
$d\sigma^A_{\mathrm{q}}/d^2\vec{k}_\perp$ represents
{\it monopole} elastic scattering\footnote{This point
is discussed in detail in~\cite{JKT}.}, in contradistinction with the
dipole scattering cross section \eq{dsigmaqqdk} appearing in DIS.

Comparing \eq{dsigmaDISdkstruct} and~\eq{dsigmaDYdkstruct} 
we interpret the non-universality of 
${d\sigma^A}/{d^2\vec{k}_\perp}$ in our model as a direct consequence of the type of 
object which interacts with the nuclear target. In the $r_\perp$-integral giving 
${d\sigma_{DIS}^A}/{d^2\vec{k}_\perp}$, the dipole 
wave function $\psi$ selects $r_\perp \lesssim 1/m_\pl$ in 
${d\sigma_{\qq}^A}/{d^2\vec{k}_\perp}$. Therefore, the hard 
scale\footnote{In the symmetric kinematics \eq{symmetric} we have 
$m_\pl^2 = y\,Q^2 +m^2 \simeq y\,Q^2$.} $m_\pl$ 
enters the physics of rescattering and is expected to play a major 
role for $k_\perp$ broadening. Conversely, in \eq{dsigmaDYdkstruct} the scale 
$m_\pl$ enters ${d\sigma_{DY}^A}/{d^2\vec{k}_\perp}$ only 
through a target-independent factor and is therefore not expected to govern 
$k_\perp$ broadening.  

In order to display the differences between DIS and DY 
we investigate the ratio
\beq
R(k_\perp) = \frac{1}{A} \left. \frac{d\sigma^A}{d^2\vec{k}_\perp} \right/ 
\frac{d\sigma^p}{d^2\vec{k}_\perp} \ \ ,
\label{Rratio}
\eeq
where $\sigma^p$ and $\sigma^A$ denote the DIS or DY cross sections
off a single scattering centre and on $A$ centres, to leading
order in $g^2$ and next-to-leading order in the target
thickness\footnote{Note that the limit of small target thickness $T
\propto A^{1/3}$ is consistent with the total coherence limit $R_A \ll
l_c$ (see~\eq{fullcoherence}), in which our DIS and DY
amplitudes~\eq{MtildeADIS} and~\eq{MtildeADY} have been derived.}
$T=A/S$. By expanding~\eq{dsigmaDISdkave2} and~\eq{dsigmaDYdkave2} 
we get, for $\lambda\ll k_\perp\ll m_\pl$ and keeping only the 
leading logarithms:
\beqa
R_{DIS}(k_\perp) &=& 1 - \frac{4g^4T}{5\pi m_\pl^2}
\log \frac{m_\pl}{k_\perp}+ \morder{T^2}  
\label{DISnuclearratio} \\
R_{DY}(k_\perp) &=& 1 + \frac{2g^4T}{\pi k_\perp^2}
\log \frac{k_\perp}{\lambda}+ \morder{T^2}.
\label{DYnuclearratio}
\eeqa

The latter results explicitly demonstrate the non-universality of the $k_\perp$-dependent 
distributions which was already apparent when comparing the full expressions 
\eq{dsigmaDISdkstruct} and \eq{dsigmaDYdkstruct}. As compared to the production on a single 
centre, and in the region under consideration $\lambda\ll k_\perp\ll m_\pl$, 
${d\sigma^A}/{d^2\vec{k}_\perp}$ is slightly reduced in DIS and strongly enhanced in DY 
production. In the DY case, we recall that for a single scattering centre (see section 3.2), 
the $k_\perp$ distribution for $k_\perp \gg \lambda$ equals that at Born level, \ie, in 
the absence of rescattering (see Eq.~\eq{dsigmaDYdklimit}). This might have suggested 
$k_\perp$-broadening to be reduced in DY as compared to DIS. The explicit calculation with 
a finite size target of `thickness' $T$ shows that this does not happen: the deviation 
from unity of the ratio $R_{DY}(k_\perp)$ survives the $\lambda \to 0$ limit.  

It is simple to realize that the $k_\perp$-broadening defined as 
\beq
\Delta \ave{k_\perp^2} \equiv \ave{k_\perp^2}_A - \ave{k_\perp^2}_p \, ,
\eeq
although different in DIS and DY, scales as $g^4T$ both in DIS and DY. Eva\-lua\-ting $\Delta \ave{k_\perp^2}$ from 
\eq{DISnuclearratio} and \eq{DYnuclearratio} (and using $d\sigma^p/d^2\vec{k}_\perp \propto 1/k_\perp^2$)
puts some light on the difference between DIS and DY. In DIS the small probability 
$\sim T/m_\pl^2$ for the rescattering of a dipole of size $\sim 1/ m_\pl$ is compensated by 
a large typical momentum transfer $k_\perp^2 \sim m_\pl^2$. In DY the hard scale $m_\pl$ 
does not enter the expression of $R(k_\perp)$, and the relatively large (monopole) rescattering probability
$\sim T/k_\perp^2$ is now compensated by a small typical transfer $k_\perp^2 \ll m_\pl^2$.
As already discussed, the non-universality of nuclear $k_\perp$-broadening in the coherent limit studied in
our model is a natural consequence of the type of object (dipole or monopole) which interacts with the target.

\section{Summary and outlook}\label{se:summary}

Within an explicit scalar QED model, we have studied transverse momentum distributions
in the coherent limit $x\ll 1$ for DIS and the DY process. 
In the aligned jet kinematics, where the leading quark or the DY pair 
carries most of the projectile momentum, the distribution in the
transverse momentum $K_\perp$ of these particles is universal, both 
for pointlike and extended targets. This is consistent with the 
universality of the $K_\perp$-dependent target quark distribution. 
On the contrary, in the symmetric kinematical region, the relevant
transverse momentum $k_\perp$ ($k_\perp \neq K_\perp$) 
is that of the hard subsystem, \ie, of 
the quark-antiquark pair of the photon fluctuation in the case of DIS, 
and of the DY virtual photon and the final antiquark in the case of
DY. The transverse momentum transfer $k_\perp$ between the target and the hard subsystem
is different in DIS and DY already for pointlike targets. The extension
to a finite size `nuclear' target stresses the physical origin of this difference. 
In DIS the $\qq$ dipole rescatters with a small probability but undergoes large 
$k_\perp$-kicks, whereas the DY $k_\perp$ distribution is sensitive to {\it monopole} 
rescattering, more likely but involving smaller kicks. 

We stress that our nuclear transverse momentum distributions are expressed in 
terms of the dipole cross section $\sigma_{\qq}$ and of the thickness
function $T$, and contain factors of the type $1-e^{-T \sigma}$ which are expected  
for classical scattering. This behaviour is a consequence of the statistical 
average we have performed on the positions of the scattering centres.  
Thus also in the coherent region rescattering off an extended target 
turns out to have the nature of stochastic multiple scattering \cite{GlauberM}. 
Coherence is important in that it fixes the nature of the object (dipole or monopole) that 
rescatters.

The non-universality of the $k_\perp$ distribution found in our scalar QED model in the coherent 
small $x$ limit calls for a systematic study of the nuclear $k_\perp$-broadening 
measured presently in fixed-target experiments ($x \sim 0.1$) or in 
the near future at RHIC ($x \sim 0.01$). 
Indeed, at the present stage we cannot give a quantitative answer to the puzzling observation 
mentioned in the Introduction, namely the smallness of transverse momentum broadening in 
DY production, as compared for instance to the broadening of the dijet momentum 
imbalance in dijet photoproduction. 
We however emphasize that with our notations the observed transverse momentum is 
$q_\perp = K_\perp$ in DY production instead of $k_\perp$ in dijet photoproduction (analogous to our DIS 
process in the symmetric kinematics). We have shown that 
$d \sigma_{DIS}/ d^2\vec{k}_\perp \neq d \sigma_{DY}/ d^2\vec{k}_\perp$, thus it is even more natural to expect   
the distributions $d \sigma_{DIS}/ d^2\vec{k}_\perp$ and $d \sigma_{DY}/ d^2\vec{q}_\perp$ with respect to {\it distinct}
transverse momentum variables to be different in the coherent regime. This can be actually checked 
explicitly in our SQED model. The universal ratio 
\beq
R(K_\perp) = \frac{1}{A} \left. \frac{d\sigma^A}{d^2\vec{K}_\perp} \right/ 
\frac{d\sigma^p}{d^2\vec{K}_\perp} 
\label{RratioK}
\eeq
can be obtained from \eq{dsigmadKave2} and \eq{dsigmadKsingle}, and reads at leading order in $g^2$ and for 
$K_\perp \ll m_\pl$:
\beq
R(K_\perp) = R_{DY}(q_\perp) = R_{DIS}(p_{1\perp}) = 
1+ \frac{2 g^4T}{\pi m_\pl^2} \log^2 \left( \frac{m_\pl}{\lambda} \right) + \morder{T^2} \,.
\label{Knuclearratio}
\eeq
We see that contrary to the `dijet' leptoproduction ratio $R_{DIS}(k_\perp)$ given in \eq{DISnuclearratio}, 
the ratio $R_{DY}(q_\perp \ll m_\pl)$ exceeds unity. Small $q_\perp$'s are favoured in DY production off a nucleus.
Whether the latter result, obtained in our abelian model, can explain the observed smallness of $q_\perp$-broadening 
in the hadronic world will be addressed in a future work.

\vskip 1cm 
{\bf Acknowledgements.} We thank P.~Hoyer and A.~Smilga for interesting discussions and useful comments. 
\vskip 1cm 

\appendix
\centerline{\Large \bf Appendix}

\section{DY production with composite projectile}

\begin{figure}[b]
\begin{center}
\epsfig{figure=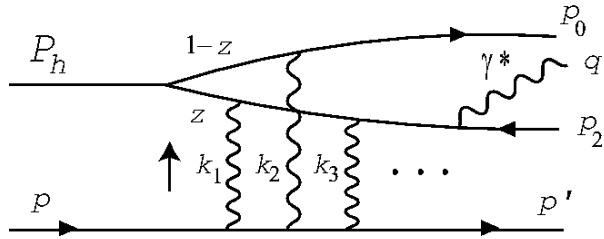,angle=0,width=8cm}
\end{center}
\caption[*]{Model for DY production with a composite projectile. 
The spectator is produced with final momentum $p_0$.}
\label{compfig}
\end{figure}

Here we study the role of spectators in DY production. 
For this purpose we extend the model of Fig.~1b to the case of a
composite projectile. The corresponding model is depicted in
Fig.~\ref{compfig}. 

The fluctuation of the `hadron' projectile of mass $m_h$ and 
momentum $P_h$ into the spectator 
quark (of final momentum $p_0$) and the active antiquark (of final
momentum $p_2$), is described by a scalar cubic coupling $ig_0$. 
The projectile is chosen to carry the `electromagnetic' charge $e$ but
to be neutral with respect to the `strong' interaction of coupling
$g$. The spectator quark has no electric charge but strong charge
$g$. The large incoming light-cone momentum $P_h^-$ splits
into the active antiquark and spectator momenta with finite fractions 
$z$ and $1-z$. The different momenta appearing in Fig.~3 
read (we choose $m_h < 2m$ in order to forbid the 
$h \rightarrow \qu \qbm$ decay):
\beqa
&& P_h = (\frac{m_h^2}{2\nu} , 2\nu, \vec{0}_\perp) \ \ ;\ \ 
p = (M, M, \vec{0}_\perp) \nonumber \\
&& p_0 = (\frac{p_{0\perp}^2+m^2}{(1-z)2\nu}, (1-z)2\nu,
\vec{p_0}_\perp) \ \ ;\ \ 
p_2 = (\frac{p_{2\perp}^2+m^2}{p_2^-}, p_2^-, \vec{p_2}_\perp)
\nonumber \\ 
&& q = (\frac{Q^2+q_{\perp}^2}{z 2\nu}, z 2\nu, \vec{q}_\perp) \simeq
(Mx_B, z 2\nu, \vec{q}_\perp) \, ,
\label{momenta}
\eeqa
where now $x_B = Q^2/(z 2M\nu)$.
We will use again the limit $x_B \ll 1$, as well as the 
kinematics defined in~\eq{scales}. 

The DY virtual photon can be radiated either by the projectile or
by the active antiquark, both having electric charge $e$. However,
in the kinematics \eq{scales}, the typical times associated to
the fluctuations $h \rightarrow \qu \qbm$ and 
$\qbm \rightarrow \gamma^* \qbm$ are respectively of order 
$\nu/p_{0\perp}^2 \to \infty$ and $1/Mx_B$. Thus the photon is radiated 
{\it after} the $h \rightarrow \qu \qbm$ fluctuation. The diagrams
where the virtual photon is emitted from the projectile are suppressed 
(in Feynman gauge) by a factor 
$\sim p_{0\perp}^2/(\nu Mx_B) \sim \morder{p_2^-/\nu}$ according to \eq{scales}.

\subsection{Consistency with factorization}

The covariant calculation of the DY production amplitude of
Fig.~\ref{compfig} is similar to the calculations performed in
Refs.~\cite{bhmps,peigne}. 
The leading-twist contribution is obtained for a virtual photon with 
longitudinal polarization 
\beq
\epsilon_L = (\frac{Q}{z \nu}, -\frac{Q}{z \nu}, \vec{0}_\perp) \, .
\label{pol}
\eeq
Going to transverse coordinate space,
\beq
{\M}(\rvec_\perp, \Rvec_\perp, \vec{u}_\perp) = \int
\frac{d^2\pvec_{2\perp} d^2\kvec_\perp d^2\pvec_{0\perp}}{(2\pi)^6}\,
\M(\pvec_{2\perp},\kvec_\perp, \pvec_{0\perp})\,
e^{i(\rvec_\perp\cdot\pvec_{2\perp} + \Rvec_\perp\cdot\kvec_{\perp} +
  \vec{u}_\perp\cdot\pvec_{0\perp})}  
\label{Fourier2}
\eeq
and resumming Coulomb scatterings yields the result
\beq
{\M}_{DY}(\rvec_\perp, \Rvec_\perp, \vec{u}_\perp) =
{\M}_{DY}(\rvec_\perp, \Rvec_\perp) \, 
e^{-ig^2 G(|\vec{R}_{\perp}+\vec{u}_{\perp}|)}\, \frac{\phi(z, u_{\perp})}{z} \, .
\label{phase2}
\eeq
Here ${\M}_{DY}(\rvec_\perp, \Rvec_\perp)$ is the DY
production amplitude in the absence of spectator obtained from \eq{normamp} 
and \eq{phase}, the function $G$ is defined in \eq{G} and $\phi(z, u_{\perp})$ is the 
$h \rightarrow \qu \qbm$ wave function
\beqa
\phi(z, u_{\perp}) &=& g_0 \, z(1-z) \, V(\delta u_{\perp})
\label{phi} \\
\delta^2 &=& m^2 - z(1-z) m_h^2  
\label{deltasquared}
\eeqa
which can be represented as
\beqa
\phi(z, u_{\perp}) &=& \int \frac{d^2\pvec_{0\perp}}{(2\pi)^2}\,
\phi(z, p_{0\perp}) \, e^{i\vec{u}_\perp\cdot\pvec_{0\perp}} 
\label{phifourier}
\\
\phi(z, p_{0\perp})&=& g_0 \,\frac{z(1-z)}{p_{0\perp}^2 + \delta^2} \, .
\eeqa

The phase factor in~\eq{phase2} arises from Coulomb rescatterings
of the spectator quark. Those indeed contribute to the DY production
amplitude of Fig.~\ref{compfig}, since vanishingly small light-cone energies 
$k_i^+ \propto 1/\nu$ can be transferred to the
spectator without any cost. The finite energy 
$k^+ = \sum k_i^+ \sim \morder{Mx_B}$ is transferred to
the active antiquark in order to produce the final state 
invariant mass $\sim Q^2$. 
The phase in~\eq{phase2} is infrared divergent (the
spectator carries the charge $g$) but this divergence compensates that
appearing in~\eq{phase}, as expected for dipole rescattering 
\beq
{\M}_{DY}(\rvec_\perp, \Rvec_\perp, \vec{u}_\perp) =
- \, e^{ig^2 W(\vec{u}_{\perp}, \vec{R}_{\perp})} \, \frac{\phi(z, u_{\perp})}{z}
\, {\M}_{DIS}(\rvec_\perp, \Rvec_\perp) \, ,
\label{phase3}
\eeq
where we used
$G(R_{\perp})-G(|\vec{R}_{\perp}+\vec{u}_{\perp}|)=W(\vec{u}_{\perp},
\vec{R}_{\perp})$.

We now proceed as in section 2 
(see Eq.~\eq{dsigma0} and following). In the presence
of the spectator the differential DY cross section is of the 
form\footnote{For the purposes of the present Appendix we do not need to 
specify the normalization of differential cross sections in the following.}
\beq
\frac{d\sigma_{DY}}{d^2\vec{q}_\perp} \propto \int
\frac{d^2\pvec_{2\perp}}{(2\pi)^2}\, \frac{d^2\pvec_{0\perp}}{(2\pi)^2}\,
|\M_{DY}(\pvec_{2\perp},
\vec{k}_{\perp}=\pvec_{2\perp}+\pvec_{0\perp}+\vec{q}_{\perp},
\pvec_{0\perp})|^2  \, .
\eeq   
Going to transverse coordinate space and using \eq{phase2} leads to
\beqa
\frac{d\sigma_{DY}}{d^2\vec{K}_\perp} &\propto& \int d^2\vec{r}_{\perp}
d^2\vec{R}_{\perp}d^2\vec{u}_{\perp}d^2\vec{r}^{\ '}_{\perp}
d^2\vec{R}^{\,'}_{\perp} d^2\vec{u}^{\,'}_{\perp} \, 
\phi(z, u_\perp) \phi(z, u'_\perp)^* \nonumber \\
&& \times \ 
\delta^2(\vec{r}_{\perp}+\vec{R}_{\perp}-
\vec{r}^{\ '}_{\perp}-\vec{R}^{\ '}_{\perp}) \,
\delta^2(\vec{u}_{\perp}+\vec{R}_{\perp}-
\vec{u}^{\ '}_{\perp}-\vec{R}^{\ '}_{\perp})  \nonumber \\
&& \times \ 
e^{ig^2(G(R_{\perp})-G(R^{\ '}_{\perp}))} \,
e^{-i(G(|\vec{R}_{\perp}+\vec{u}_{\perp}|)-G(|\vec{R}^{\
    '}_{\perp}+\vec{u}^{\ '}_{\perp}|))}  
\nonumber \\
&& \times \ 
e^{-i(\vec{R}_{\perp}-\vec{R}^{\ '}_{\perp})\cdot \vec{q}_{\perp}}\,
{\M}_{DIS}(\vec{r}_{\perp}, \vec{R}_{\perp})
{\M}_{DIS}^*(\vec{r}^{\ '}_{\perp}, \vec{R}^{\ '}_{\perp}) \, .
\label{dsigmadq}
\eeqa
From the $\delta$-constraints the Coulomb phase associated to
spectator rescattering cancels out, 
$G(|\vec{R}_{\perp}+\vec{u}_{\perp}|)-G(|\vec{R}^{\ '}_{\perp}+
\vec{u}^{\ '}_{\perp}|) \to 0$, and the remaining phase difference 
$G(R_{\perp})-G(R^{\ '}_{\perp})$ is absorbed in the expression 
of ${\M}_{DIS}$ given by \eq{MtildeDIS} and \eq{dipamp}, leading to
\beqa
\frac{d\sigma_{DY}}{d^2\vec{q}_\perp} &\propto& \int d^2\vec{r}_{\perp}
d^2\vec{R}_{\perp}d^2\vec{r}^{\ '}_{\perp}d^2\vec{R}^{\ '}_{\perp}
\, \delta^2(\vec{r}_{\perp}+\vec{R}_{\perp}-
\vec{r}^{\ '}_{\perp}-\vec{R}^{\ '}_{\perp}) \nonumber \\
&& \times \ e^{-i(\vec{R}^{\ '}_{\perp}-\vec{R}_{\perp})\cdot \vec{q}_{\perp}}
\ {\M}_{DIS}(\vec{r}_{\perp}, \vec{R}_{\perp})
{\M}_{DIS}^*(\vec{r}^{\ '}_{\perp}, \vec{R}^{\ '}_{\perp})
\nonumber \\
&& \times \ \int d^2\vec{u}_{\perp} \phi(z, u_\perp) 
\phi(z, |\vec{u}_\perp+\vec{R}_\perp-\vec{R}^{\ '}_{\perp}|)^*  \, .
\label{dsigmadq2}
\eeqa
Thus spectator Coulomb rescattering does not affect 
$d\sigma_{DY}/d^2\vec{q}_\perp$ (and a fortiori not the total
leading-twist DY cross section either), which is consistent with 
factorization, as we briefly see now.

From~\eq{phifourier} one gets
\beq
\int d^2\vec{u}_{\perp} \phi(z, u_\perp) 
\phi(z, |\vec{u}_\perp+\vec{R}_\perp-\vec{R}^{\ '}_{\perp}|)^* =
 \int \frac{d^2\pvec_{0\perp}}{(2\pi)^2}\,
|\phi(z, p_{0\perp})|^2 \, 
e^{-i(\vec{R}^{\ '}_\perp-\vec{R}_\perp)\cdot\pvec_{0\perp}}
\eeq
Inserting this into~\eq{dsigmadq2} and identifying 
$\int d^2\pvec_{0\perp} |\psi(z, p_{0\perp})|^2$ with the projectile
antiquark distribution 
$f_{\qbm/h}(z, -\vec{p}_{0\perp})$ we obtain
\beq
\frac{d\sigma_{DY}}{d^2\vec{q}_\perp} \propto \int d^2\pvec_{0\perp}
f_{\qbm/h}(z, -\vec{p}_{0\perp}) 
f_{\qu/T}(x_B, \vec{p}_{0\perp}+\vec{q}_{\perp})  \, .
\label{convolution}
\eeq
The latter equation shows that our DY model with spectator is consistent 
with factorization theorems involving $K_\perp$-dependent parton 
distributions \cite{fact2}. 

\subsection{Non-universality of $k_\perp$ Coulomb exchange}

Here we argue that the result found in section 3.2, namely that the typical 
$k_\perp$ contributing to 
$\Delta\sigma_{DY} = \sigma^{\rm tot}_{DY} - \sigma^{\rm Born}_{DY}$ is 
$k_\perp \sim \lambda \to 0$, naturally translates to 
$k_\perp \sim \delta$ in the case of a composite projectile of size 
$R_h \sim 1/\delta$.    
 
The $k_\perp$-distribution reads
\beqa
\frac{d\sigma_{DY}}{d^2\vec{k}_\perp} &\propto& \int
\frac{d^2\pvec_{2\perp}}{(2\pi)^2}\, \frac{d^2\pvec_{0\perp}}{(2\pi)^2}\,
|\M_{DY}(\pvec_{2\perp},\vec{k}_{\perp},\pvec_{0\perp})|^2  \nonumber
\\
&\propto& \int d^2\vec{u}_{\perp} |\phi(z, u_\perp)|^2 \int d^2\vec{r}_{\perp}
d^2\vec{R}_{\perp}d^2\vec{R}^{\ '}_{\perp} \, 
e^{-i(\vec{R}_{\perp}-\vec{R}^{\ '}_{\perp})\cdot \vec{k}_{\perp}} 
\nonumber \\
&& \ \times \ \ e^{ig^2 (W(\vec{u}_{\perp}, \vec{R}_{\perp}) -
  W(\vec{u}_{\perp}, \vec{R}^{\ '}_{\perp}))} 
\, {\M}_{DIS}(\vec{r}_{\perp}, \vec{R}_{\perp})
{\M}_{DIS}^*(\vec{r}_{\perp}, \vec{R}^{\ '}_{\perp}) \, ,
\nonumber \\
\label{dsigmaDYdkt}
\eeqa
where we used~\eq{phase3}.
The integrand of~\eq{dsigmaDYdkt} depends on the scales $m_\pl$ and 
$\delta$, the latter corresponding (see~\eq{phi}) to the inverse 
transverse size of the projectile. 
When $\delta \to 0$, $u_\perp \sim 1/\delta \to \infty$, and the 
(finite) Coulomb phase in~\eq{dsigmaDYdkt} becomes
\beqa
W(\vec{u}_{\perp}, \vec{R}_{\perp})-W(\vec{u}_{\perp}, 
\vec{R}^{\ '}_{\perp}) &=& G(R_{\perp}) - G(R^{\ '}_{\perp}) + \frac{1}{2\pi} 
\log\left(\frac{|\vec{u}_{\perp}+\vec{R}_{\perp}|}{|\vec{u}_{\perp}+
\vec{R}^{\ '}_{\perp}|}\right)  \nonumber \\
&& \hskip -1cm \mathop{\longrightarrow}_{u_\perp \to \infty} \ \ 
G(R_{\perp}) - G(R^{\ '}_{\perp}) 
\eeqa
We thus recover, in the $\delta \to 0$ limit, the
$k_\perp$-distribution \eq{dsigmaDYdkt0} 
in the DY model without spectator, for which we have shown that 
$k_\perp \sim \lambda \to 0$. In other words, when the size of the
projectile 
$R_h \sim 1/\delta \to \infty$, the spectator plays no screening role
any longer. In practice $\delta$ is non-zero, $\delta \sim m$ (but still 
$\delta \ll m_\pl \simeq  \sqrt{y} \, Q$ in the kinematical region
\eq{symmetric}) and the typical $k_\perp$ contributing to 
$\Delta\sigma_{DY}$ is of order $\delta$, the largest infrared cut-off 
at disposal.

\end{document}